\documentclass[aps,pre,amsmath,lengthcheck,superscriptaddress]{revtex4-2}

\usepackage{graphicx}\graphicspath{ {figures/} }
\usepackage{hyperref}
\usepackage{amssymb}
\hypersetup{colorlinks,allcolors=blue,breaklinks}

\newcommand{\be}{\begin{equation}}
\newcommand{\ee}{\end{equation}}
\newcommand{\ba}{\begin{align}}
\newcommand{\ea}{\end{align}}
\newcommand{\bi}{\begin{itemize}}
\newcommand{\ei}{\end{itemize}}

\newcommand{\bla}{bla\\bla\\bla\\bla\\bla}

\begin{document}

\title{Fluctuation-optimization theorem}

\author{Pierre Naz\'e}
\email{pierre.naze@unesp.br}

\affiliation{\it Universidade Federal do Par\'a, Faculdade de F\'isica, ICEN,
Av. Augusto Correa, 1, Guam\'a, 66075-110, Bel\'em, Par\'a, Brazil}

\date{\today}

\begin{abstract}

A fluctuation theorem relating the work to its optimal average work is presented. The function mediating the relation is increasing and convex, and depends on the switching time $\tau$, driving strength $\delta\lambda/\lambda_0$, and protocol $g(t)$. The result is corroborated by an example of an overdamped white noise Brownian motion subjected to a moving laser harmonic trap. Observing also that the fluctuation-optimization theorem is an Euler-Lagrange equation, I conclude that the function minimizing $\langle h(-\beta W)\rangle$ obeys the relation proposed. The optimal work can now be calculated with numerical methods without knowing the optimal protocol, using only a work distribution of an arbitrary protocol.

\end{abstract}

\maketitle

\section{Introduction} 

The determination of the optimal work performed in a thermodynamic process has become a central problem in recent times~\cite{deffner2020thermodynamic}. This is typically addressed by using analytical or numerical methods to derive the optimal protocol, which its use in the work functional leads to this desired minimum value. Some problems may exist in such a procedure, like the unavailability of numerical methods to achieve distribution contributions in the optimal work~\cite{naze2023analytical}. Effective methods to avoid such problems become therefore important.

In this work, I start to develop an alternative idea to find the optimal work: by using statistical samples of a work performed with an arbitrary protocol, I calculate via a fluctuation theorem the optimal work performed employing an appropriate function that mediates the equality. I am going to show that such a function is globally increasing and convex, deriving then the optimization condition, such as Jarzynski's equality reproduces the Second Law of Thermodynamics~\cite{jarzynski1997nonequilibrium}. An example is presented at the end corroborating the derived results. Also, observing that such fluctuation-optimization theorem is an Euler-Lagrange equation associated with an extreme of a convex functional, I conclude that the desired function minimizes the relation $\langle h(-\beta W)\rangle$. In this manner, numerical methods, such as genetic programming, could effectively find such a function by only knowing a work distribution.

\section{Fluctuation-optimization theorem} 

Consider a thermodynamic driven system, thermally isolated or not, with a Hamiltonian $\mathcal{H}$, depending on an external parameter $\lambda(t)=\lambda_0+g(t)\delta\lambda$. During the switching time $\tau$, the protocol $g(t)$ is changed from $g(0)=0$ to $g(\tau)=1$. The system is prepared in a thermal state of temperature $\beta^{-1}$. The average work performed on the system for several repetitions of this process is
\be
\langle W\rangle(\tau)=\int_0^\tau \langle\partial_\lambda \mathcal{H}\rangle(t)\dot{\lambda}(t)dt
\ee
I suppose that there exists an optimal protocol $g^*(t)$ such that the optimal average work $\langle W\rangle^*$ calculated under its driving obeys the optimality condition
\be
\langle W\rangle\ge \langle W\rangle^*,
\label{eq:optimalinequality}
\ee
for all possible protocols in a certain switching time $\tau$. The aim is to find an increasing and convex function $f(x)$ such that
\be
\langle f(-\beta W)\rangle=f(-\beta \langle W\rangle^*).
\label{eq:fot}
\ee
If such a function exists, according to Jensen's inequality and properties of increasing functions it holds Eq.~\eqref{eq:optimalinequality}. 

Consider then a number $w$ in the support of $W$. Expanding $f(-\beta W)$ in a Taylor's series around $x=w$, one has
\be
f(-\beta W) = \sum_{n=0}^{\infty}\frac{f^{(n)}(w)}{n!}(-\beta(W-w))^n.
\ee
Expanding now $f(-\beta \langle W\rangle^*)$ in a Taylor's series around $x=w$, one has
\be
f(-\beta \langle W\rangle^*) = \sum_{n=0}^{\infty}\frac{f^{(n)}(w)}{n!}(-\beta(\langle W\rangle^*-w))^n.
\ee
To satisfy the fluctuation theorem, one should have
\be
\sum_{n=0}^{\infty}\frac{f^{(n)}(w)}{n!}(-\beta)^n(\langle (W-w)^n\rangle -(\langle W\rangle^*-w)^n) =0.
\ee
Considering Jarzynski's equality, the proposed fluctuation theorem is satisfied if one chooses
\be
f^{(n)}(w)=\frac{\langle (W-w)^n\rangle -(\Delta F-w)^n}{\langle (W-w)^n\rangle -(\langle W\rangle^*-w)^n},
\ee
where $\Delta F$ is Helmholtz's free energy variation between the final and initial equilibrium states. Let us analyze the important cases now. For $n=0$, I take the limit $n\rightarrow 0$ using L'Hôspital's rule
\be
f^{(0)}(w)=\frac{\langle (W-w)^{-1}\rangle -(\Delta F-w)^{-1}}{\langle (W-w)^{-1}\rangle -(\langle W\rangle^*-w)^{-1}}.
\ee
For $n=1$, one has
\be
f^{(1)}(w)=\frac{\langle W\rangle -\Delta F}{\langle W\rangle -\langle W\rangle^*}\ge 0,
\ee
since $\langle W\rangle- \Delta F\ge 0$ according to Jarzynski's equality and $\langle W\rangle- \langle W\rangle^*\ge 0$ for the optimality condition. As the result holds for all $w$ in the support of $W$, then $f(x)$ is a global increasing function. For $n=2$, one has
\be
f^{(2)}(w)=\frac{\langle (W-w)^{2}\rangle -(\Delta F-w)^{2}}{\langle (W-w)^{2}\rangle -(\langle W\rangle^*-w)^{2}}\ge 0,
\ee
since $h(x)=(x-w)^2$ is a convex function, one can use Jensen's inequality to show that the denominator and numerator are greater than zero. As the result holds for all $w$ in the support of $W$, $f(x)$ is a global convex function. Therefore, the fluctuation theorem proposed derives the optimality condition. Indeed, since the function is convex, by using Jensen's inequality in Eq.~\eqref{eq:fot}, one has
\be
f(-\beta \langle W \rangle)\le  f(-\beta\langle W\rangle^*).
\ee
Using now the increasing condition, one derives
\be
\langle W \rangle\ge  \langle W\rangle^*.
\ee
Remark that the coefficients $f^{(n)}(w)$ highly depend on the switching time $\tau$, the driving strength $\delta\lambda/\lambda_0$ and the protocol $g(t)$. Also, using the lower bound $\Delta F$ for all possible optimal work $\langle W\rangle^*$ in this fluctuation-optimization theorem, one recovers Jarzynski's equality.

\section{Example}

\begin{figure}[t]
    \centering
    \includegraphics[scale=0.7]{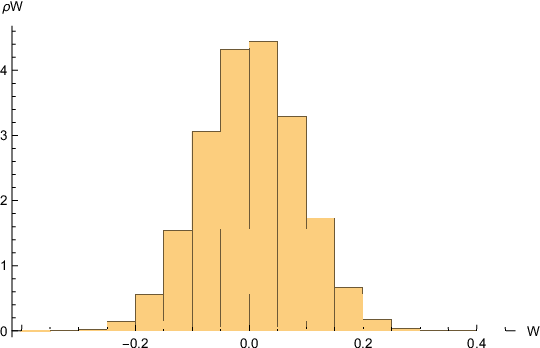}
    \caption{Numerial work distribution calculated for an overdamped white noise Brownian motion subject to a moving laser harmonic trap. It was used a unit mass, $\beta=1$, $\omega_0=1$, $\gamma=1$, $\delta\lambda/\lambda_0=0.1$, $\tau=\gamma/\omega_0^2$.}
    \label{fig:1}
\end{figure}

Consider an overdamped white noise Brownian motion subject to a moving laser harmonic trap, governed by the following Langevin equation
\be
\dot{x}+\frac{\omega_0^2}{\gamma}(x(t)-\lambda(t))=\eta(t),
\ee
where $\eta(t)$ is a white noise. Here, I consider a unit mass, a damping coefficient $\gamma$, and natural frequency $\omega_0$. The driven parameter $\lambda(t)$ is the equilibrium position, given by
\be
\lambda(t)=\lambda_0+\delta\lambda t/\tau.
\ee
Considering $\beta=1$, $\gamma=1$, $\omega_0=1$, $\lambda_0=1$ and $\delta\lambda=0.1$, I sample $10^6$ initial conditions from the canonical distribution to numerically calculate the work performed under such a linear driving. I consider also $\tau=\gamma/\omega_0^2$ and a time step $\Delta t=0.01$. The work distribution is given by Fig.~\ref{fig:1}. Observe that the support of $W$ is concentrated around $w=0$. 

\begin{figure}[t]
    \centering
    \includegraphics[scale=0.7]{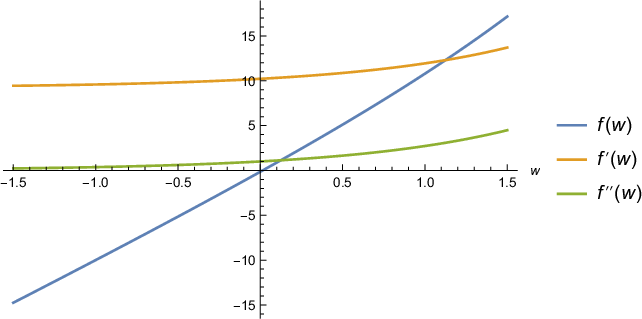}
    \caption{Function $f(x)$, its first derivative $f'(x)$ and its second derivative $f''(x)$.}
    \label{fig:2}
\end{figure}

To calculate $f(x)$, I consider the Helmholtz's free energy variation between the final and initial equilibrium states, and the optimal work given by~\cite{naze2022optimal}
\be
\Delta F = 0,
\ee
\be
\langle W\rangle^* = \frac{\omega_0^2\delta\lambda^2}{2+\omega_0^2\tau/\gamma}.
\ee
 Considering Taylor's series expansion around $w=0$ until $10$th order, the function $f$, its first and second derivatives are illustrated in Fig.~\ref{fig:2}. Indeed, the derivatives are positive in a considerable region around $w=0$, corroborating the fluctuation-optimization theorem. Also, one has
\be
\langle f(-\beta W)\rangle-f(-\beta \langle W\rangle^*)\approx 1.44\times 10^{-5},
\ee
which corroborates the result as well.

\section{Fluctuation-optimization theorem as Euler-Lagrange equation}

Let us show that the fluctuation-optimization theorem is nothing more than an Euler-Lagrange equation associated with the extremum of a convex functional. Also, let us see that the function $f(x)$ can be recovered by the minimization of $\langle h(-\beta W)\rangle$.

Consider first that the fluctuation theorem demonstrated in this work is easily generalized for any $W'$, with $\Delta F\le W'\le \langle W\rangle^*$. Considering the following convex functional $S[h]$
\be
S[h] = (\langle h(-\beta W) \rangle - h(-\beta W'))^2,
\ee
I observe that the Euler-Lagrange equation associated is the following fluctuation theorem
\be
\langle h^*(-\beta W) \rangle = h^*(-\beta W'),
\ee
where $h^*$ is increasing and convex. Since $S$ is convex, it presents a minimum. Therefore, one has
\be
S[h^*]\le S[h].
\ee
From such a relation, it holds
\begin{widetext}
\begin{equation}
\begin{split}
((\langle h(-\beta W) \rangle+\langle h^*(-\beta W) \rangle)&-(h(-\beta W')+h^*(-\beta W')))\times\\
&((\langle h(-\beta W) \rangle-\langle h^*(-\beta W) \rangle)-(h(-\beta W')-h^*(-\beta W')))\ge 0
\end{split}
\label{eq:ssign}
\end{equation}
\end{widetext}
From such a relation, every factor should have the same sign. Assuming non-negative, one has
\be
\langle h(-\beta W) \rangle+\langle h^*(-\beta W) \rangle\ge h(-\beta W')+h^*(-\beta W'),
\label{eq:prod1}
\ee
\be
\langle h(-\beta W) \rangle-\langle h^*(-\beta W) \rangle\ge h^*(-\beta W')-h(-\beta W').
\label{eq:prod2}
\ee
Summing up Eqs.~\eqref{eq:prod1} and ~\eqref{eq:prod2}, one has
\be
h^*(-\beta W')\le\langle h(-\beta W) \rangle.
\ee
The fluctuation-optimization theorem holds as well
\be
\langle h^*(-\beta W)\rangle\le\langle h(-\beta W) \rangle.
\ee
As suggested by this lower bound, let us find that $h^*$ and $W'$ to make the following relation true
\be
h^*(-\beta W')=\underset{h}{\text{min}}\langle h(-\beta W) \rangle.
\ee
Observe that $h^*(x)$ is increasing, so $h^{*(-1)}(x)$ is as well. Therefore
\be
W'=\underset{h}{\text{max}}\left[-\frac{1}{\beta}h^{*(-1)}(\langle h(-\beta W) \rangle)\right],
\ee
whose maximum value is achieved for $W'=\langle W\rangle^*$ with $h^*(x)=f(x)$. Assuming now from the relation~\eqref{eq:ssign} that the factors have non-positive signs, following the same reasoning used before, one has
\be
\langle h^*(-\beta W)\rangle\ge\langle h(-\beta W) \rangle,
\ee
with
\be
W'=\underset{h}{\text{min}}\left[-\frac{1}{\beta}h^{*(-1)}(\langle h(-\beta W) \rangle)\right],
\ee
whose minimum value is achieved for $W'=\Delta F$ and $h^*(x)=\exp(x)$. Thus, the convex functional $S[h]$ has two fluctuation theorems inside it under the extremization of $\langle h(-\beta W)\rangle$: Jarzynski's equality and the fluctuation-optimization theorem.

Therefore, $f(x)$ is such that its average is the minimum of all possible averages taken with other functions. Thus, finding a numerical method that calculates the minimum of this average will furnish the function $f$ necessary to calculate the average work without knowing the optimal protocol. For example, genetic programming~\cite{koza1994genetic} would be a nice first attempt to verify the fluctuation theorem. Observe that the unique information needed is the work distribution, calculated with an arbitrary protocol.

\section{Conclusion} 

The fluctuation-optimization theorem was derived in this work. The function that mediates the relation is globally increasing and convex and derives the optimal condition such as Jarzynski's equation derives the Second Law of Thermodynamics. I presented an example that corroborates the derived results. Also, I conclude that techniques such as genetic programming may be useful to find such function, since it minimizes the relation $\langle h(-\beta W)\rangle$. In this way, the optimal work can be calculated without knowing the optimal protocol.

\bibliography{FOT.bib}
\bibliographystyle{apsrev4-2}

\end{document}